\documentclass[aip,rsi,preprint,superscriptaddress, showpacs, endfloat]{revtex4}

\usepackage{bm}        
\usepackage{color}
\usepackage{gensymb}
\usepackage{epstopdf}

\usepackage{graphicx,amsmath,epsfig}

\begin{document}


\title{An in-situ method for measuring the non-linear response of a Fabry-Perot cavity}


\author{Wenhao Bu, Mengke Liu, Dizhou Xie, Bo Yan}


\email{yanbohang@zju.edu.cn}
\affiliation{Department of Physics, Zhejiang University, Hangzhou, Zhejiang, China, 310027}


\date{\today}

\begin{abstract}
High finesse Fabry-Perot(FP) cavity is a very important frequency reference for laser stabilization, and is widely used for applications such as precision measurement, laser cooling of ions or molecules. But the non-linear response of the piezoelectric ceramic transducer (PZT) in the FP cavity limits the performance of the laser stabilization. Measuring and controlling such non-linearity are important. Here we report an in-situ, optical method to characterize this non-linearity by measuring the resonance signals of a dual-frequency laser. The differential measurement makes it insensitive to laser and cavity drifting, and has a very high sensitivity. It can be applied for various applications with PZT, especially in an optical lab.
\end{abstract}


\maketitle

\section{Introduction}
Stabilization of the frequency of a laser is a very important technique in modern physics, it has lots of applications such as optical clock \cite{cole:2013,Bloom:2014,zhang:2014}, gravitational wave detection \cite{ligo}, laser cooling of atoms, molecules \cite{Shuman:SrF2010,Yeo:2015} or ions and ultracold molecule creation \cite{yan2013:dipole-dipole,Moses:2015}. For laser cooling technique, the lasers are usually required to be stabilized to MHz-level which is narrower compare with the nature linewidth of excited states of atoms, molecules or ions. In atomic case, such as alkali atoms, the atomic transition can be easily monitored in a vapor cell and used as references to lock lasers. But for molecules or ions, it is hard to monitor the transitions. In these cases, transfer cavity is a widely-used, low-cost and easy-setup method \cite{transfer_cavity:1994,transfer_cavity2,Bohlouli2006}.

In order to use the transfer cavity locking method, the cavity needs to be locked to a stable reference first, such as a He-Ne laser. This can be done by monitor the transmission (or reflection) signal while scanning the voltage applied to the piezoelectric ceramic transducer (PZT) of the cavity. The peak (dip) signal indicates the resonant frequency of the cavity, and gives the information of the cavity length. A bias voltage is applied and varied to make sure the resonant position doesn't shift, thus the cavity is locked \cite{Barry:2013,Dai:2014}. Then the target laser can be locked to a different position, the laser frequency is set by,

\begin{equation}
f_1=f_0+f(V).
\end{equation}
where $f_0$ depends on the cavity length without applying a voltage to the PZT, and $f(V)$ is the voltage-frequency function of the cavity. It depends on the length change of the PZT,
\begin{equation}
f(V)=\frac{\Delta L(V)}{\lambda/4}f_{FSR}
\end{equation}
for a confocal cavity (for the confocal cavity the definition of free spectrum range (FSR) is different than the normal cavity, here $f_{FSR}=c/4d$, $d$ is the cavity length). For the first order perturbation, the length changes linearly with the applied voltage, $\Delta L(V)=\kappa \Delta V$, which is widely used \cite{Barry:2013,Yin:2015}. But for a PZT, the non-linearity always happens. If the bias voltage changes a lot, the error could be serious and affect the laser stabilization. It worths to measure and character such effect. This can be done by using the precise machine to measure the displacement or stress\cite{PZT1,PZT2}. But this machine is not common in an optical lab, and it is usually hard to take the PZT out for a commercial FP cavity. So an in-situ measurement will be very convenient and useful.

The voltage-frequency function of the PZT response can be measured by using formula (1). Varying the laser frequency and monitoring the resonant position, we can get a plot of such function. But this direct measurement needs a stable laser ($f$ stable) and a stable cavity ($f_0$ stable). The shift of either laser or cavity will effect the final result. That is not an easy requirement, both laser and the cavity can slowly drift due to the changes of the temperature, pressure or other noise.

\begin{figure}[t]
\includegraphics[width=5in]{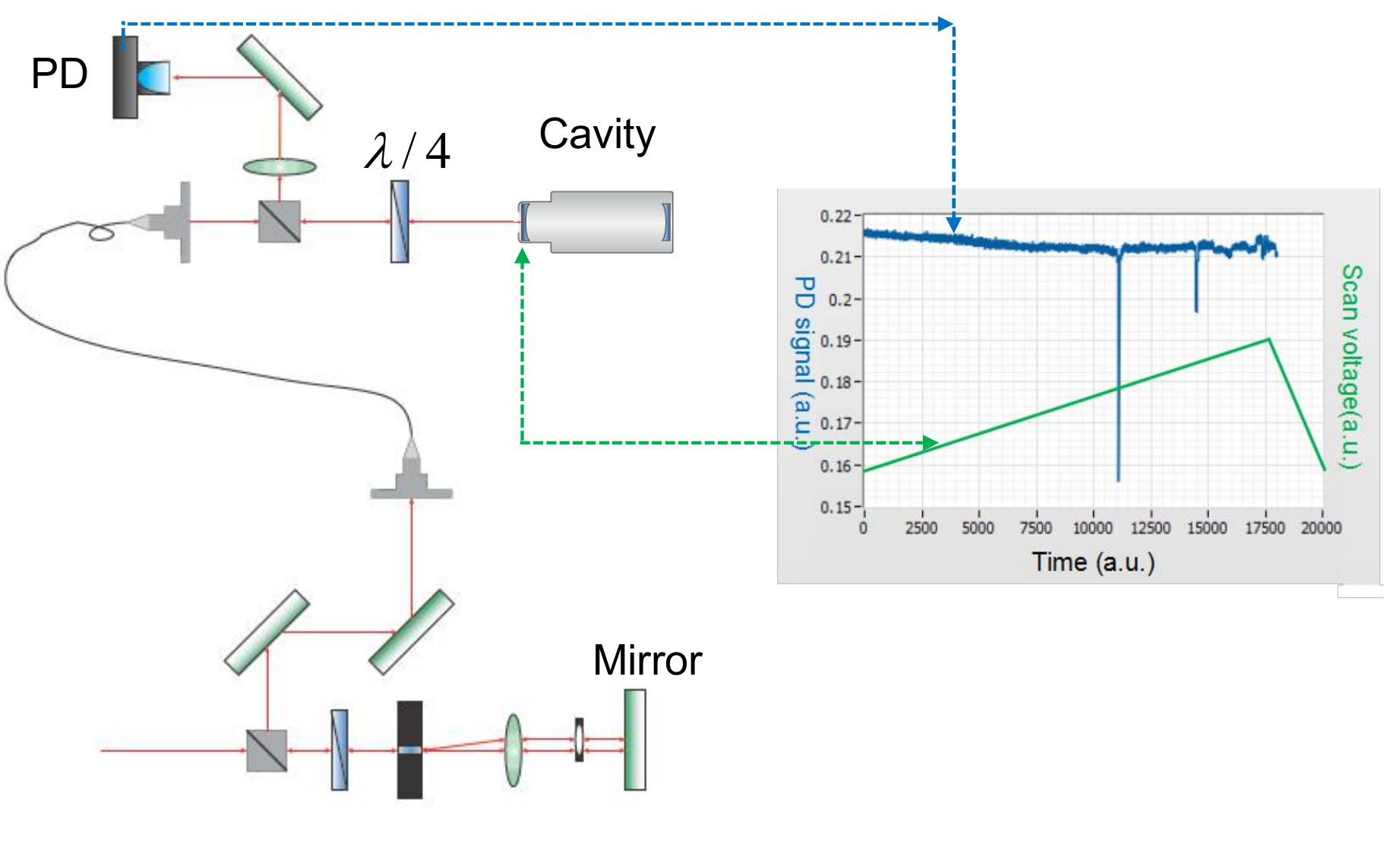}
\caption[setup]{(Color online) the experimental setup. One laser beam double pass the AOM and then couple to the fiber. Both zero order and two-time first order beams share the same beam path and send to the cavity. A scanning voltage is applied to the PZT of the cavity. The reflection signal is record by the photon detector (PD). The typical data shows two resonant points which is due to the resonance of dual-frequency laser.}\label{fig:setup}
\end{figure}

In order to solve such problem, we develop a differential method. Instead of sending the laser with one frequency, we send a laser with two frequencies (with fixed frequency difference $\Delta f$) to the cavity. Double peaks (dips) will show up in the photon detector (PD). From the two resonant points, the voltage difference $\Delta V$ can be traced. In this way, we get the differential data,
\begin{equation}
\frac{df}{dV}\approx \frac{\Delta f}{\Delta V}.
\end{equation}
By taking the differential measurement, the driftings of the laser and the cavity become the common noises and get canceled.
A set of differential data can be measured with different bias voltages. So the response function can be reconstructed,
\begin{equation}
 f(V)=\int \frac{df}{dV} dV \approx \sum_i (\frac{\Delta f}{\Delta V})_i(\Delta V)_i.
\end{equation}

\section{Experimental setup}
Figure 1 shows the experimental setup. A laser beam double pass an acoustic-optical modulator (AOM), and then couple to the fiber. The double-pass setup makes the zero order and the two-time $+1$ order share the almost the same beam path \cite{double-pass}, and both couple to the fiber. In this way, the dual-frequency laser is ready and send to the FP cavity, the reflected light is collected by a PD, and recorded by a computer. The cavity used in the experiment is a commercial confocal cavity (Topitica FPI 100). The finesse is about $>430$, The FSR is 1G, and about 9V per FSR for 893nm laser. The typical data shows two resonant points when scanning the PZT. The position difference tell the voltage difference $\Delta V$.  We average about 100 shots for each point. Then decrease the bias voltage to shot next FSR, repeat this process, we get a plot for $df/dV$ with the voltage $V$ as shown in Fig.2 (a).

\section{Data analysis}

Figure 2(a) shows the differential data for both $898nm$ and $860nm$ lasers. Both cases show the clear slopes which mean non-linearity of the PZT. It can be very well fit with a linear relation. So it is better to assume the frequency-voltage relation is
\begin{equation}
f(V)=\alpha V+\beta V^2.
\end{equation}

Since we measure the frequency response instead of length response with V, $f(V)$ dependents on the wavelength as indicated by formula (2). As a double check, we define a ratio between two lasers,
\begin{equation}
\eta=\frac{(df/dV)_{898}}{(df/dV)_{860}}\times \frac{898nm}{860nm}.
\end{equation}
Insert of Fig. 2(a) shows the ratio, the different between this ratio and unit is less than $1\%$, consists with the theory.

\begin{figure}[t]
 \includegraphics[width=3.4in]{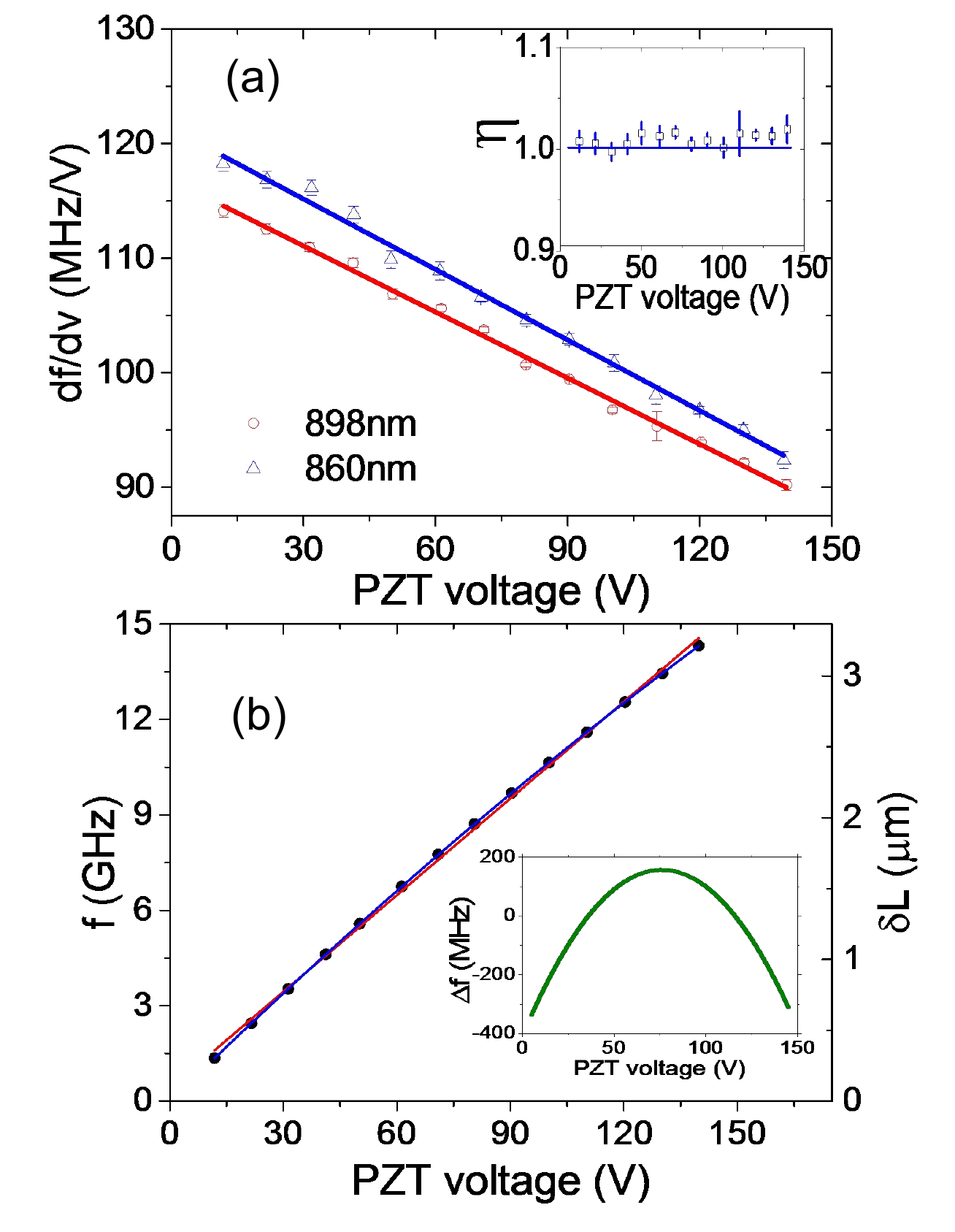}%
 \caption[data]{(Color online) The experimental data. (a) the differential data verse voltage. Both 898nm and 860nm laser shows a clear slope, which is due to the non-linearity of PZT. The insert shows the length ratio measured by two different laser. (b) reconstruction of the frequency-voltage function. The data points (black dot) are calculated from (a) with formula (4). The red line is the linear fitting plot, the blue curve is the parabola fitting. The insert shows the difference between the two fittings. }
\label{fig:App}
 \end{figure}

In Fig. 2(b), we reconstruct the frequency-voltage relation using formula (4). we fit the data with both linear and parabola function. The parabola function fits much better, and we plot the difference between two fitting function in the insert. The data shows if the bias changes few tens volts, the accumulated error for laser frequency could be few hundred MHz, which is too big for laser cooling technique.

\section{Conclusion}
In this paper, we report a differential method to measure the non-linear response of the PZT in a FP cavity. The method can not only be used to character the PZT response, but also for in-situ and real time correction for laser stabilization. In this way, the hystersis can be compensated and the long term lock of the laser can be much better. Noise of the laser and the drifting of the cavity is canceled, thus achieve a high sensitivity with a simple setup. All the experimental components are very common in an optical lab, this method is very suitable for an optical lab, and should be able to applied to various PZT applications.

\begin{acknowledgments}
This work is supported by ¡°the Fundamental Research Funds for the Central Universities 2016QNA3007¡±.

\end{acknowledgments}

\bibliographystyle{aipauth4-1}
\bibliography{PZT-bib}
\end{document}